\documentclass[aps,preprint,prd,epsfig]{revtex4}

\usepackage[english]{babel}
\usepackage[utf8x]{inputenc}
\usepackage[T1]{fontenc}

\usepackage[a4paper,top=3cm,bottom=2cm,left=2cm,right=2cm,marginparwidth=1.75cm]{geometry}

\usepackage{amsmath}
\usepackage{graphicx}
\usepackage[colorinlistoftodos]{todonotes}
\usepackage[colorlinks=true, allcolors=blue]{hyperref}
\begin{document}
\title{ Resonant production of $Z^{\prime}$  and signature of right-handed neutrinos  within a 3-3-1 model}
\author{F. F. Freitas, C. A. de S. Pires, P. Vasconcelos}
\affiliation{{Institute of Theoretical Physics, Chinese Academy of Sciences,
 Zhong Guan Cun East Street 55, P. O. Box 2735, Beijing 100190, P. R. China},\\
 { Departamento de Física, Universidade Federal da Paraíba, Caixa Postal 5008, 58051-970,
Jo\~ao Pessoa, PB, Brazil}}

\begin{abstract}
 Recent simulation studies indicate that right-handed neutrinos may be discovered  at future high luminosity LHC runs as long as a new neutral gauge boson, $Z^{\prime}$, has been discovered first. In this case, when $Z^{\prime}$  is resonantly produced at the LHC it, subsequently, decays into pair of right-handed neutrinos with mass up to TeV  scale. Then the decay of these neutrinos leave as signature  tri-lepton final states and/or same-sign di-muon and a boost di-boson. The prospect of discovering right-handed neutrinos through this mechanism has been developed within abelian gauge extension of the standard model. In this work we study the processes involved in this discovery mechanism within the framework of the 3-3-1 model
 \end{abstract}
\maketitle
\section{INTRODUCTION}
The future runs of the LHC  at tens of TeVs and High luminosity  lead  to the tantalizing possibility of probing the TeV Type-I seesaw mechanism for small neutrino masses\cite{Yanagida:1979as,GellMann:1980vs,Glashow:1979nm,Mohapatra:1979ia}. The signature of this mechanism is right-handed neutrino (RHN) with  mass around TeV or less. The challenge with the production of such RHNs depends on their  mixing with the standard neutrinos which is very tiny,  $\mathcal{O}(10^{−6}$). The tiny mixing diminishes considerably their production at the LHC leaving us with very few events which difficult we have a  good statistical significance.

To circumvent this problem, it was recently proposed that the discovery of RHNs may be conditioned to the existence of a new massive neutral gauge boson, $Z^{\prime}$, which interacts with all fermions, including the RHNs\cite{Kang:2015uoc,Cox:2017eme,Accomando:2017qcs}. Then it is considered that $Z^{\prime}$ is resonantly produced at the LHC through the process $pp \rightarrow Z^{\prime}$, and that subsequently  it decays in a pair of RHNs,  $Z^{\prime} \rightarrow SS$. Numerical simulations, having as benchmark model the $B-L$ gauge model, have showed  that under certain specific conditions, the signature of these neutrinos may be detected at the future Large hadron Collider (LHC) if they are kinematically accessible at the LHC.  The signature of RHN in this case is trilepton final
state $\ell^\pm \ell^\mp \ell^\mp \nu_\ell jj$, or same-sign dimuon and a boosted diboson $\ell^\mp\ell^\mp W^\pm W^\pm$. The authors of
Ref.\,\cite{Cox:2017eme}  considered the latter channel and concluded that , for fixed masses, $m_{Z^\prime} = 3$
TeV and $m_S = m_{Z^\prime} /4$, if we have $\sigma(pp \rightarrow Z^\prime \rightarrow SS \rightarrow \ell^\mp\ell^\mp W^\pm W^\pm) \approx 0.1$, we may have a 5$\sigma$ discovery at the LHC if we have a luminosity of $300 \text{fb}^{−1}$. For the trilepton final states, a signal-to-background  ratio of $S/\sqrt{B} \simeq 10$ requires $\sigma(pp\rightarrow Z^{\prime} \rightarrow S_3 S_3 \rightarrow \ell^\pm \ell^\mp \ell^\pm \nu_\ell jj) \simeq 0.37$fb for a luminosity of 300fb$^{-1}$ with $m_{Z^{\prime}}=4$TeV and $m_{S_3}=400$GeV, see   Ref.\,\cite{Accomando:2017qcs}.

In this paper we study the behavior of all processes involved in this mechanism  within the $SU(3)_C \otimes SU(3)_L \otimes U(1)_N$  gauge models with right-handed neutrinos (331$\nu_R$)\cite{Montero:1992jk,Foot:1994ym}. The idea  is to verify if the model fulfills the conditions that may lead to the discovery of RHNs. This model performs the type-I seesaw mechanism in a very peculiar way since the mass matrix of the RHNs and of the standard ones, both, involves the same Yukawa couplings\cite{Cogollo:2008zc,Dong:2008sw,Ky:2005yq}. This guarantee that the heaviest RHN has mass of 1TeV or less, which are accessible kinematically at the future LHC. Its particle content has also a new neutral gauge boson $Z^{\prime}$. The main aim is to check if the model provides $\sigma(pp \rightarrow Z^\prime \rightarrow SS \rightarrow \ell^\mp\ell^\mp W^\pm W^\pm)$ and $\sigma(pp \rightarrow Z^\prime \rightarrow SS \rightarrow \ell^\mp\ell^\mp  \ell^\pm jj\nu)$ around $10^{-1}$\,fb at the resonance of $Z^{\prime}$ for the LHC running at 14\,TeV with 300\,fb$^{-1}$.

This work is organized as follows. In Sec. (II) we present the main aspect of the model,  develop the seesaw mechanism and derive the neutral and charged current of the neutrinos with the gauge bosons of the model. In Sec. (III) we discuss resonant production of $Z^{\prime}$ and its subsequent decay in pairs of RHNs and show in graphics the behavior of the cross sections involved in the mechanism with $m_{Z^{\prime}}$ and $m_S$. Finally, in Sec. (IV), we present our conclusions.

\section{ THE MECHANISM}
\subsection{Revisiting the model}

In this section we revisit the TeV type I seesaw mechanism within  the 331$\nu_R$ \cite{Cogollo:2008zc}. Before, we present the main aspect of the  model\cite{Montero:1992jk,Foot:1994ym}.

The leptonic content is arranged in triplet and singlet:
\begin{equation}
f_{aL}= \begin{pmatrix}
\nu_{a}     \\
\ell_{a}       \\
\nu^{c}_{a} \\
\end{pmatrix}_{L} \sim (1,3,-1/3), \quad l_{aR}\sim (1,1,-1),
\end{equation}
where $\sim$ represent the way the triplet transform under the 3-3-1 symmetry. The index $a(=1,2,3)$ represents the three known generations of leptons. Notice that the third component of the lepton triplet is the right-handed neutrino.

In the quark sector, the first and second generations transform as  anti-triplet, while the third generation transform as triplet under $SU(3)_L$ symmetry
\begin{equation}
Q_{iL}=\left( 
\begin{array}{c}
d_i \\
-u_i\\
d^{\prime}_i \\
\end{array}\right)_{L}\sim  (3, \bar{3},0), i=1,2,\quad   
Q_{3L}= \left( 
\begin{array}{c}
u_3 \\
d_3 \\
u^{\prime}_3 \\
\end{array}\right)_L\sim (3, 3, 1/3).
\end{equation}
In addition to the left-handed field, we have right-handed fields as singlets under 3-3-1 symmetry
\begin{eqnarray}
&& u_{iR}\sim (3,1,2/3),\quad u_{3R}\sim (3,1,2/3), \quad u^{\prime}_{3R}\sim (3,1,2/3),  \nonumber \\
&& d_{iR}\sim (3,1,-1/3),\quad d_{3R}\sim (3,1,-1/3),\quad d^{\prime}_{iR}\sim (3,1,-1/3),
\end{eqnarray} 
with $i=1,2$.  

The original scalar sector of the model involves three scalar triplets
\begin{equation}
\eta=\left(\begin{array}{c}
\eta^0\\
\eta^-\\
\eta^{\prime 0}\\
\end{array}\right),\quad
\rho=\left(\begin{array}{c}
\rho^+\\
\rho^0\\
\rho^{\prime +}\\
\end{array}\right),\quad
\chi=\left(\begin{array}{c}
\chi^0\\
\chi^{-}\\
\chi^{\prime 0}\\
\end{array}\right),
\end{equation}
with $\eta$ and $\chi$  transforming as $(1,3,-1/3)$ and $\rho$ transforming as  $(1,3,2/3)$  under 3-3-1 symmetry. This content is sufficient to generate the correct masses for all massive fields \cite{Montero:1992jk,Foot:1994ym, Hoang:1995vq}, except neutrinos. We remember here that, in addition to the standard gauge bosons, $Z_\mu, W^{\pm}_\mu$ and the photon, $A_\mu$, the model has five new massive gauge bosons $Z^{\prime}_\mu$, $W^{\prime \pm}_\mu$, $U^0_\mu$, $U^{0 \dagger}_\mu$. 
\subsection{The TeV type-I seesaw mechanism}
In order to generate neutrino masses, we add a scalar sextet to the original scalar content \cite{Foot:1994ym,Ky:2005yq,Dong:2008sw}:
\begin{equation}
\label{SextetoEscalar}
\ensuremath{\mathcal{S}}=\frac{1}{\sqrt{2}}\left(\begin{array}{ccc}
\Delta^{0} & \Delta^{-} & \Phi^{0} \\
\newline \\
\Delta^{-} & \Delta^{--} & \Phi^{-} \\
\newline \\
\Phi^{0} & \Phi^{-} & \sigma^{0} \end{array}\right) \sim (1,6,-2/3).
\end{equation}
It was showed in \cite{Cogollo:2008zc}  that with this scalar sextet  TeV type-I seesaw mechanism is easily implemented. In what follow we revisit such result. The potential of the model with sextet is developed in \cite{Cogollo:2008zc} .



After the spontaneous  breaking of the  $SU(3)_C \times SU(3)_L \times U(1)_N$ to the standard symmetry,  $SU(3)_C \times SU(2)_L \times U(1)_Y$, the scalar sextet splits into a triplet, a doublet and a singlet of scalar,
\begin{equation}
S \rightarrow \Delta_{(1,3,Y_\Delta )} + \Phi_{(1,2,Y_\Phi)} + \sigma_{(1,1,Y_\sigma)}
\end{equation}
where $Y_\Delta = −2$, $Y_\Phi = −1$ and $Y_\sigma = 0$ are the hyper-charges of the respective multiplets with
\begin{equation}
\label{multipletos}
\Delta = \frac{1}{\sqrt{2}}\left(\begin{array}{cc}
\Delta^{0} & \Delta^{-} \\
\newline \\
\Delta^{-} & \Delta^{--} \end{array}\right), \quad 
\Phi = \frac{1}{\sqrt{2}}\left(\begin{array}{c}
\Phi^{0} \\
\newline \\
\Phi^{-} \\
\end{array}\right), \quad \sigma = \frac{\sigma^{0}}{\sqrt{2}}.
\end{equation}

The Yukawa interaction formed by the leptonic triplets and the scalar sextet is
\begin{equation}
\label{LyukawaS}
G_{ab} \overline{(f_{aL})^{c}} \mathcal{S}^{*}(f_{bL}).
\end{equation}
This interaction is factorized into the following ones
\begin{equation}
\label{}
\overline{(f_{aL})^{c}} \mathcal{S}^{*}(f_{bL}) \rightarrow \overline{(L_{aL})^{c}} \Delta^{*}L_{bL}  + \overline{L_{aL}} \Phi^{*}(\nu_{bR}) + \overline{(\nu_{aR})^{c}} \mathcal{\sigma}^{0*}(\nu_{bR}).
\end{equation}
with $$L_a = \left(\begin{array}{c}
\nu_a \\
\ell_a \\
\end{array}\right)_L .$$

Assuming $\langle\Delta\rangle = 0$, $\langle\Phi\rangle = \frac{v_\Phi}{\sqrt{2}}$ and $\langle\sigma^{0}\rangle = \frac{v_\sigma}{\sqrt{2}}$, the Yukawa interactions in Eq. (\ref{LyukawaS}) yields the following Dirac and Majorana mass terms for the neutrinos in  basis $\left( \nu_{eL}, \nu_{\mu L}, \nu_{\tau L}, (\nu_{eR})^{c}, (\nu_{\mu R})^{c},(\nu_{\tau R})^{c} \right)^{T} = \left(\nu_{L},(\nu_{R})^{c} \right)^T$

\begin{eqnarray}
\frac{1}{2}\left(\begin{array}{cc}
\overline{(\nu_{L})^{c}} & \overline{\nu}_{R} 
\end{array}\right)  M^{D+M} \left(\begin{array}{c}
\nu_{L}       \\
\newline      \\
(\nu_{R})^{c}
\end{array}\right),
\end{eqnarray}
where

\begin{equation}
\label{mixneutrino}
M^{D+M} = \left(\begin{array}{cc}
0 & (M_{D})^{T}        \\
\newline               \\
M_{D} & M_{R}
\end{array}\right).
\end{equation}
and
\begin{equation}
M_{D} = G v_\Phi, \quad  M_{R} = G v_\sigma,
\label{relation}
\end{equation}
with $G$ being a symmetric matrix formed by the Yukawa couplings $G_{ab}$. Throughout this paper, for simplicity, we neglect CP violation effects in the leptonic sector, which means that there are no phases in the above mass matrix and then all their elements are real. In what follow we take $v_\sigma > v_\Phi$.

The diagonalization of the mass matrix in Eq. (\ref{mixneutrino}), for the case $ v_\sigma >> v_\Phi $, leads to a mass relation among $M_D$ and $M_R$ known as the seesaw mechanism \cite{Bilenky:1998dt,GonzalezGarcia:2007ib},
\begin{equation}
\label{seesaw}
M_{\nu} \simeq  -M_D(M_R)^{-1}M_D, \text{and} \quad M_{R} 
\end{equation}
where $M_{\nu}$ is a mass matrix for the left-handed neutrinos and $M_R$ is the mass matrix for the right-handed neutrinos. 

Having said that, in the  more general case the mass matrix $M^{D+M}$ is diagonalized by the rotating matrix $U_{6\times 6}$ \cite{Bilenky:1998dt}
\begin{equation}
\label{blocodiag}
U^{T}M^{D+M}U \simeq \left(\begin{array}{cc}
M_{light} & 0       \\
\newline            \\
0 & M_{heavy} 
\end{array}\right),
\end{equation}
where $M_{light}$ is diagonal mass matrix and gives the masses of the standard neutrinos while $M_{heavy}$ is diagonal and gives the masses of the right-handed neutrinos. The rotating mix
matrix, $U$, is given by
\begin{equation}
\label{blocodiag}
U \simeq \left(\begin{array}{cc}
U_{PMNS}      & \frac{v_\Phi}{v_\sigma} U_R       \\
\newline            \\
-\frac{v_\Phi}{v_\sigma} U_{PMNS} & U_{R} 
\end{array}\right),
\end{equation}
where $U_{PMNS}$ and $U_R$ diagonalize $M_\nu$ and $M_R$, respectively. The rotating matrix $U$ connects physical eigenstates with flavor ones. From now on we adopt the following notation: the light physical neutrinos will be call $(N_1,  N_2, N_3)$, while the heavy ones will be $(S_1, S_2, S_3)$. The relations between mass
and flavor neutrinos\ eigenstates are given by:
\begin{eqnarray}
&&\nu_{L} = U_{PMNS}N_{L} + \frac{v_\Phi}{v_\sigma}U_{R}(S^{c})_{L},
\nonumber \\
&&(\nu^{c})_{L} = -\frac{v_\Phi}{v_\sigma}U_{PMNS} N_{L} + U_{R}(S^{c})_{L}, \end{eqnarray}
where $N_L = (N_1 , N_2 , N_3 )^T$ and $S_L = (S_1 , S_2 , S_3 )^T$.

Now comes one of the main points of this TeV type-I seesaw  mechanism. One should notice that, in terms of VEV’s, $M_{light} \propto \frac{v_{\Phi}^{2}}{v_\sigma}$ and $M_{heavy} \propto v_\sigma$. It is important to stress that $\sigma^0$ carries two units of lepton number. Hence, when $\sigma^0$ develops VEV, both the lepton number as well as the 3-3-1 symmetry are spontaneously broken. This means that the 3-3-1 symmetry breaking energy scale coincides with that of spontaneous breaking of the lepton number. The 3-3-1 symmetry breaking energy scale is expected to occur at TeV scale, as so will the spontaneous lepton number violation, i.e., $v_\sigma$ should belong to the TeV scale. 

On the other hand, $\Phi$ does not carry lepton number. Its VEV generates, exclusively, Dirac mass terms for the neutrinos. The proposal of the seesaw mechanism is to yield neutrino masses at eV scale, which is accomplished if the ratio $\frac{v^{2}_\Phi}{v_\sigma}$ lies around eV. However, as we argued above $v_\sigma$ should be around TeV, then the eV neutrino requires $v_\Phi$ around MeV. 

Another interesting point is that the masses of the left-handed and right-handed neutrinos both share the same  Yukawa coupling since that they originate from the same Yukawa interaction which is given in Eq. (\ref{LyukawaS}). This means that the Yukawa couplings $G_{ab}$ provides the pattern of the standard neutrinos as well as of the RHNs. This is a distinguishable fact of the model.

With the relation between the mass and flavor neutrino eigenstates, we are able to obtain the charged and neutral currents involving these neutrinos. Except by the photon, all the other gauge bosons of the model interact with the neutrinos
\begin{eqnarray}
\label{ZZprimeneutrino}
\mathcal{L}^{W} =  &-& \frac{g}{\sqrt{2}} \left[ \overline{N_{L}}(U^{T}_{PMNS})\gamma^{\mu} l_{L} + \frac{v_{\Phi}}{v_{\sigma}}\overline{S_{L}}(U^{T}_{R}) \gamma^{\mu} l_{L} \right] W_{\mu}^{+} + H.c. , \nonumber \\
\mathcal{L}^{W^{\prime}} = &-& \frac{g}{\sqrt{2}}\left[-\frac{v_\Phi}{v_\sigma}  \overline{N_{L}}(U_{PMNS}^{T})\gamma^{\mu} l_{L} + \overline{S_{L}}(U^{T}_{R}) \gamma^{\mu} l_{L} \right] W_{\mu}^{\prime +} + H.c. ,\nonumber \\
\mathcal{L}^{U} =
&-& \frac{g}{\sqrt{2}}\left[\overline{N_{L}} (U^{T}_{PMNS}) \gamma^{\mu} (U_{R}) S_{L}  - \frac{v_\Phi}{v_\sigma}\bar{N_{L}} \gamma^{\mu} N_{L} - \frac{v_\Phi}{v_\sigma}\overline{S_L} \gamma^{\mu} S_{L} \right] U^{0}_{\mu} + H.c., \nonumber \\
\mathcal{L}^{Z}=&-&\frac{g}{2C_{w}}\left[\overline{N_L}\gamma^{\mu}  N_{L} + \frac{v_\Phi}{v_\sigma}\left( \overline{N_L}(U^{T}_{PMNS}U_R)\gamma^{\mu} S_{L} + H.C. \right) \right]Z_{\mu},\\
\ensuremath{\mathcal{L}}^{Z^{\prime}}=&-&\frac{g}{2C_{W}} \frac{(1-2S_{W}^{2})}{\sqrt{3-S_{W}^{2}}}\left[\overline{N_L}\gamma^{\mu}  N_{L} + \frac{v_\Phi}{v_\sigma}\left( \overline{N_L}(U^{T}_{PMNS}U_R)\gamma^{\mu} S_{L} + H.C. \right) \right]Z^{'}_{\mu} \nonumber \\
&+&\frac{g}{2C_{W}} \frac{2C_{W}^{2}}{\sqrt{3-S_{W}^{2}}}\left[\overline{S_L}\gamma^{\mu} S_{L} - \frac{v_\Phi}{v_\sigma}\left( \overline{S_L}(U_R^{T}U_{PMNS})\gamma^{\mu} N_{L} + H.C. \right) \right]Z^{'}_{\mu},\nonumber 
\end{eqnarray}
where $l_L = (e_L , \mu_L , \tau_L )^T$ and $C_W=\cos\theta_W$ and $S_W=\sin \theta_W$ with $\theta_W$ being the Weinberg angle. For the interactions between quarks and $Z^{\prime}$, see \cite{Hoang:1995vq,Queiroz:2016gif,Cogollo:2012ek,Promberger:2007py}

The interactions of the RHNs with the standard gauge bosons, $W^\pm$ and $Z$, are suppressed by the ratio $\frac{v_\Phi}{v_\sigma}$. For typical values of $v_\Phi$ and $v_\sigma$, we obtain a suppression factor of order of $10^{−6}$ which is of the same order of magnitude that in the case of TeV type-I seesaw mechanism in $B-L$ gauge  extension of the SM. The same suppression factor appears in interactions of the standard neutrino with the 3-3-1 gauge bosons $W^{\prime}$ and $U^0$. The gauge boson $Z^{\prime}$ interacts with both standard and right-handed neutrinos while $Z$ interacts dominantly with the standard neutrinos. In other words, the RHNs interact dominantly with the 3-3-1 gauge bosons $W^{\prime }$, $U^0$ and $Z^{\prime}$. This implies RHNs are dominantly produced by the decay  channel $Z^{\prime} \rightarrow SS$. 

According to Eq. \eqref{relation} and \eqref{seesaw},  at least one RHN ($S_1$) may be light enough to have a  long life and scape the LHC detectors and be considered as missing transverse energy. In this case it will contribute to the invisible $Z$ decay, $Z \rightarrow S_1 S_1$, and we have to check if this is in agreement with  the  current experimental bound $\Gamma^{exp}_{inv}=503\pm 16$\,MeV \cite{Patrignani:2016xqp}. As it is possible that at least one standard neutrino be massless, let us assume that one RHN, $S_1$ is massless, too. According to the interactions above, the total  width decay of the standard  $Z$ will include the contributions: $Z \rightarrow \sum_{i,j} (N_i N_i + N_i S_1 + S_1S_1 )$ with $i=1,2,3$. For the  values of the  parameters considered here, and using MadGraph5 \cite{Degrande:2011ua}, we obtain $\Gamma_{inv}=507.367$\,MeV. Such prediction is in agreement with the experimental value for the invisible decay of $Z$ standard gauge boson.

Finally, the interactions between gauge bosons and other fermions of the model is given in Ref.\cite{Hoang:1995vq}. Moreover, as the scalar sector of the model is very complex, and their interactions with neutrinos are sub-dominantly, once involve Yukawa couplings, then in our analysis we discard the interactions of the RHNs with the scalars of the model.

\section{ Pair production of RHNs and their signature  }

The mass expression for the RHNs in the 331$\nu_R$ model is  $M_{heavy}=Gv_\sigma$. As the typical value of $v_\sigma$ is around TeV and the pattern of the matrix $G$ is determined by the standard neutrino data, then assuming normal hierarchy it is reasonable to expect that $G_{ij}$ take values in the range $10^{-3}-10^{-1}$, which implies that at least one RHN may have mass of hundreds of GeVs. In the APENDIX we present a realistic scenario. For our study here we consider that the heaviest RNH neutrino ($S_3$) has mass of 200\,GeV. 

Concerning RHN with mass of hundreds of GeVs, recent simulation studies considered the prospect of its discovery at future LHC\cite{Kang:2015uoc,Cox:2017eme,Accomando:2017qcs,Buchmuller:1991ce}. The framework used in the simulations was the $B-L$ model\cite{Mohapatra:1980qe,Marshak:1979fm,Wetterich:1981bx,Masiero:1982fi,Mohapatra:1982xz}. The success of the simulations is conditioned to the discovery, first, of a  new neutral gauge boson $Z^{\prime}$ with mass of few TeVs associated to $U(1)_{B-L}$. Then the idea is that when $Z^{\prime}$ is resonantly produced through the process $pp \rightarrow Z^{\prime}$, its subsequent decay in pair of RHNs, $ Z^{\prime} \rightarrow S_3\,S_3$, has as signature  tri-lepton final states, $S_3 S_3 \rightarrow l^\pm l^\pm l^\mp \nu jj$, or same-sign di-muon and a boost di-boson, $ S_3 S_3 \rightarrow \mu^\pm \mu^\pm W^\mp W^\mp$. As result, successful simulations require that   $\sigma(pp\rightarrow Z^{\prime}\rightarrow S_3 S_3 \rightarrow \mu^\pm \mu^\pm W^\mp W^\mp)$ and $\sigma(pp\rightarrow Z^{\prime}\rightarrow S_3 S_3 \rightarrow l^\pm l^\pm l^\mp \nu jj) $ take values around $0.1$\,fb for a 5$\sigma$ discovery at the LHC running at 14\,TeV with 300\,fb$^{-1}$. In this section we study the behavior of these cross sections within the 331$\nu_R$ model with the aim of checking if the signature of the RHNs of the 331$\nu_R$ model may be probed at future LHC runs.

It is interesting to inspect, first,  if the  neutrino pair production cross section surpasses the charged lepton pair production one $pp\rightarrow Z^{\prime} \rightarrow \ell^+ \ell^-$ ($\ell=e\,\,\,\,\mbox{or}\,\,\,\mu$). This is so because  $pp \rightarrow Z^{\prime} \rightarrow S_3 S_3$ competes with 
di-lepton final states  $pp\rightarrow Z^{\prime} \rightarrow \ell^+ \ell^-$. Thus, an interesting model is one that gives $\frac{\sigma(pp\rightarrow Z^{\prime} \rightarrow S_3S_3)}{\sigma( pp\rightarrow Z^{\prime} \rightarrow \ell^+ \ell^-)}>1$ in the resonance of $Z^{\prime}$.

In order to check that,  we implement the interactions of Eq. \eqref{ZZprimeneutrino} in the MadGraph5 and obtain the values for the width decays provided by the 331$\nu_R$ model.

\begin{figure}[ht]
\centering
\includegraphics[scale=0.5]{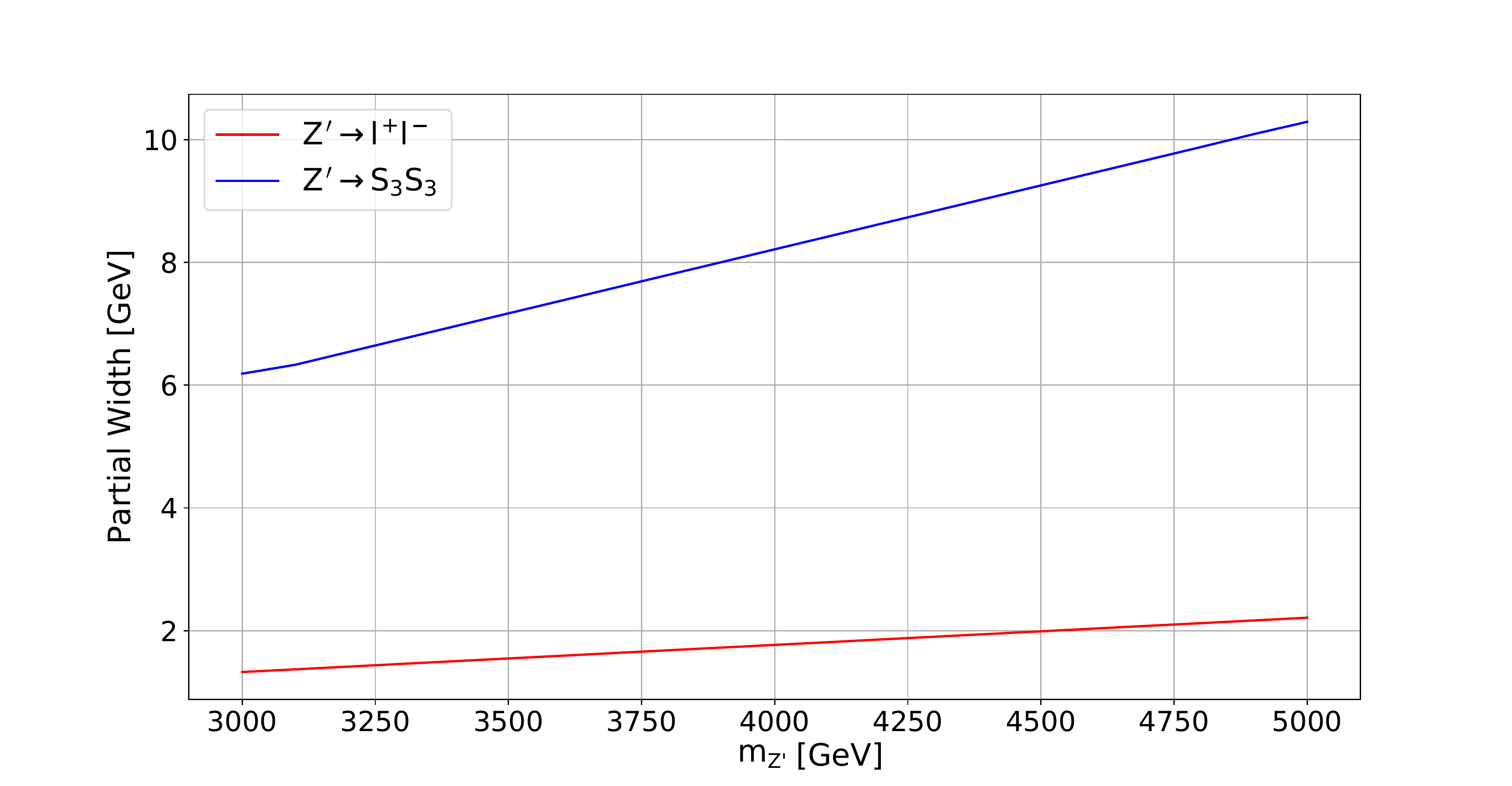}
\caption{Partial width for $\Gamma (Z^\prime \rightarrow S_3S_3 )$ (blue line)  and  $\Gamma (Z^\prime \rightarrow \ell^+ \ell^- )$(red line) varying with $m_{Z^{\prime}}$}.
\label{fig1}
\end{figure}

Such widths depend on $m_{Z^{\prime}}$ and their behavior are displayed at Figure.\eqref{fig1}. Perceive that the model provides naturally $\Gamma(Z^{\prime} \rightarrow S_3 S_3) >\Gamma(Z^{\prime} \rightarrow \ell^+ \ell^- )$ in the following rate  $\frac{\Gamma(Z^{\prime} \rightarrow S_3S_3)}{\Gamma(Z^{\prime} \rightarrow \ell^+ \ell^-)} \simeq 4.7$ (per generation). This is an encouraging result. Remember that the $B-L$ model is the benchmark framework  used in the simulations and it provides $\frac{\Gamma(Z^{\prime} \rightarrow N_R N_R)}{\Gamma(Z^{\prime} \rightarrow \ell^+ \ell^-)}=0.5$. Alternative $U(1)_X$ models may increase considerably this rate\cite{Das:2017flq,Das:2017deo,Das:2017ski}

\begin{figure}[ht]
\centering
\includegraphics[scale=0.5]{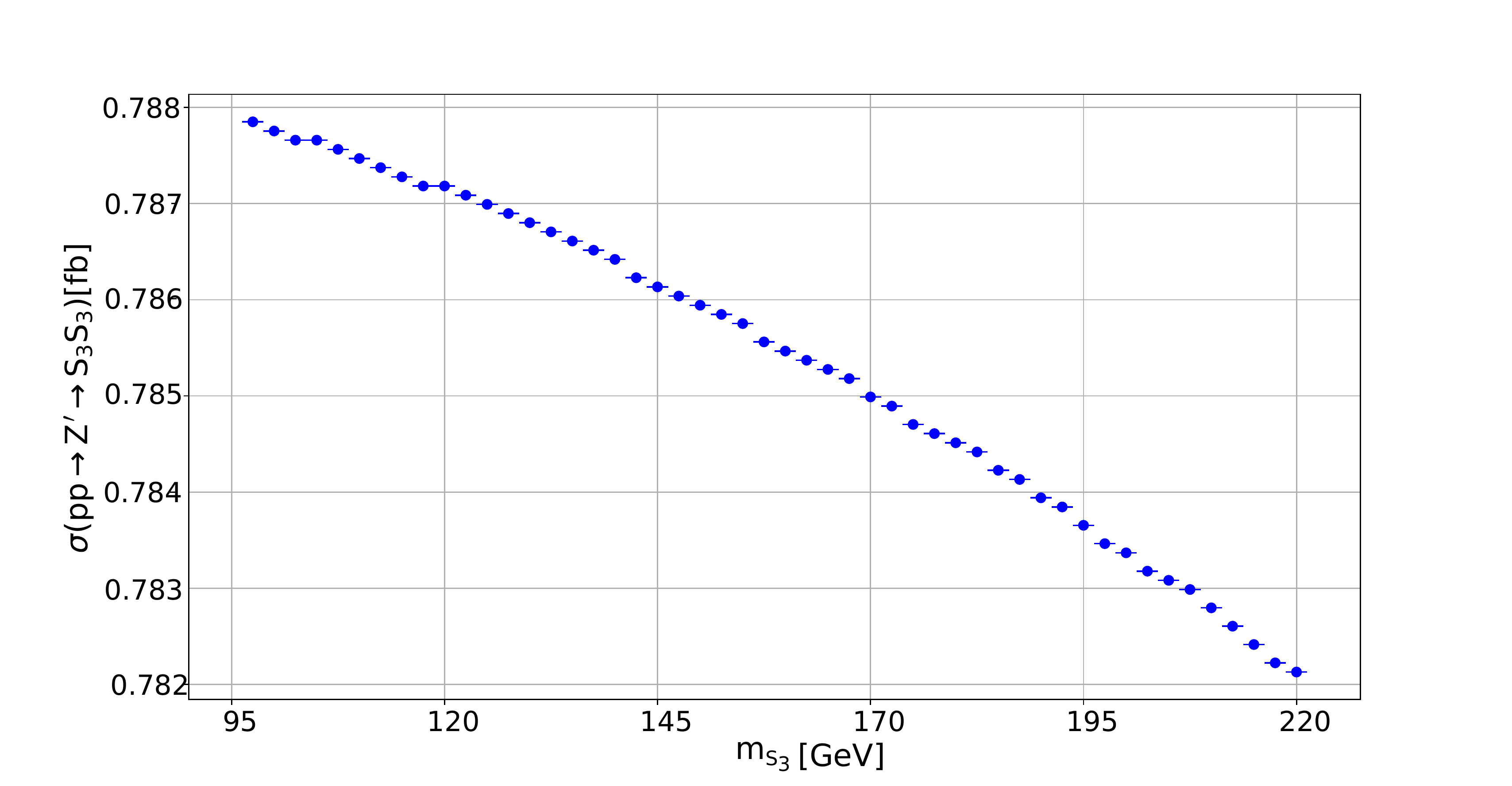}\caption{ Behavior of the cross section $\sigma( pp \rightarrow Z^{\prime} \rightarrow S_3\,S_3)$ varying with $m_{s_3}$ at LHC with 14\,TeV and $m_{Z^\prime}=4\,$TeV.}
\label{fig4}
\end{figure}

Before go further, we must examine the behavior of the  original cross section $\sigma(p\,p\rightarrow Z^{\prime} \rightarrow S_3\, S_3)$ in order to check if it falls in the range of values that allows future prospect of discovering RHNs at the LHC at high luminosity. In Figure.\eqref{fig4}, we present the behavior of this cross section with $m_{S_3}$. Note that it reproduces the RHN production cross section required for the 5$\sigma$ discovery of RHNs with 300\,fb$^{-1}$ \cite{Cox:2017eme} which is $\sigma(p\,p\rightarrow Z^{\prime} \rightarrow S_3\, S_3)\simeq 0.8$\,fb.

Now we evaluate the cross sections for tri-leptons and di-muon production within the 331$\nu_R$ model. Differently from the $B-L$ case, in the 331$\nu_R$ model the standard Higgs does not couple directly with RHNs since it is the sextet of scalars that provide the seesaw mechanism. In view of the complexity of the potential, we do not solve the scalar sector, then we neglect Higgs contributions and obtain the cross sections intermediated by the gauge bosons of the model, only.

From the interactions in Eq. (\ref{ZZprimeneutrino}), and using Madgraph5, we estimate the following branching ratios:
\begin{eqnarray}
&&BR(S_3 \rightarrow W \ell)\simeq 0.47,\quad BR(S_3 \rightarrow Z \nu_\ell)\simeq 0.35\nonumber \\
&& BR(W \rightarrow \ell\nu_\ell)\simeq 0.22,\quad BR(W \rightarrow \jmath\jmath)\simeq 0.67, \quad BR(Z \rightarrow \ell^{+}\ell^{-})\simeq 0.068, 
\label{BRs}
\end{eqnarray}
where $\ell=e\,\,\text{and}\,\,\mu$.

\begin{figure}[ht]
\centering
\includegraphics[scale=0.5]{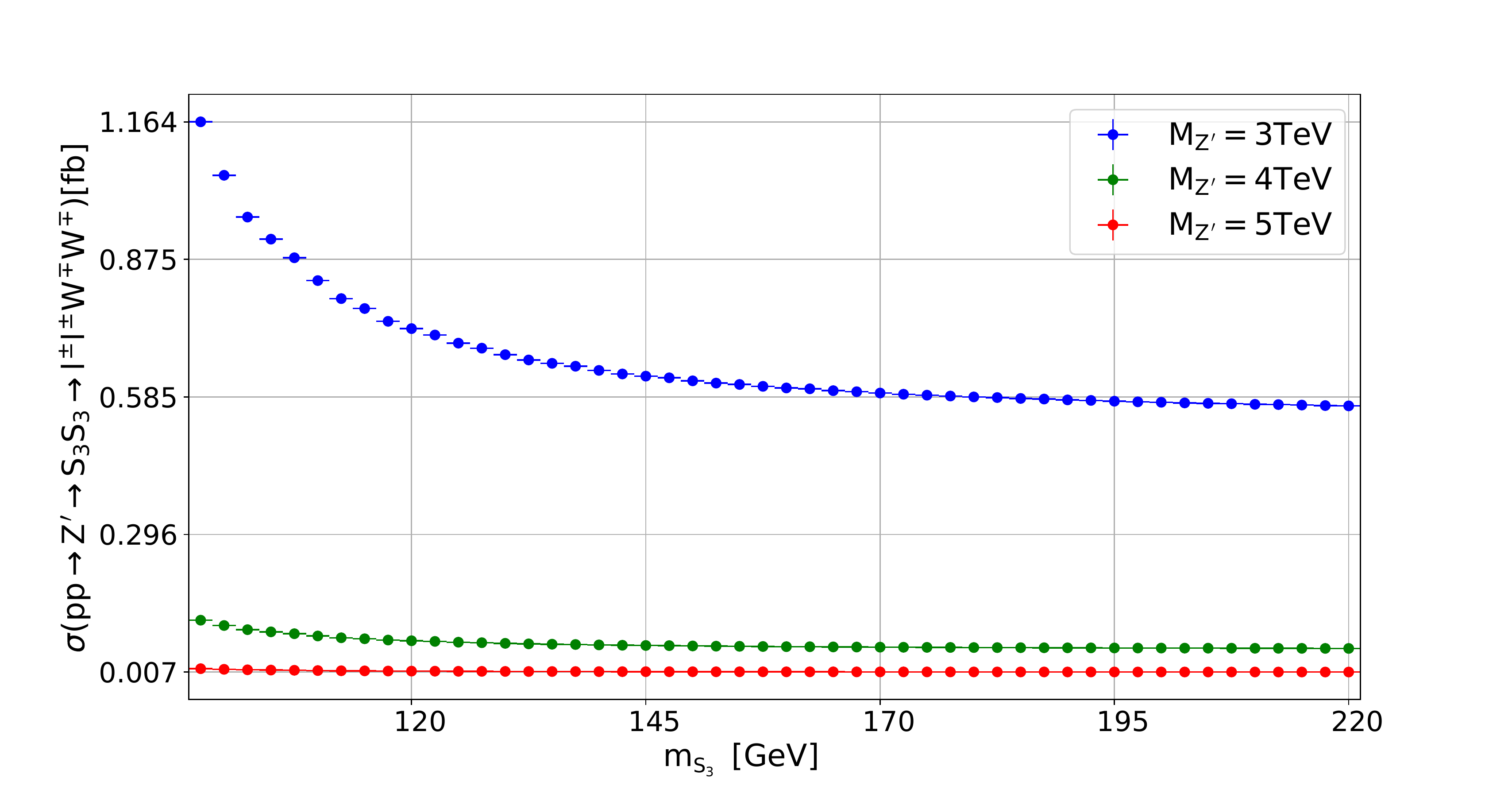}
\caption{ Behavior of the cross section $\sigma (pp \rightarrow Z^{\prime} \rightarrow S_3\,S_3 \rightarrow \ell^{\pm}\ell^{\pm}W^{\mp}W^{\mp})$ varying with $m_{s_3}$ at LHC with 14\,TeV. The colors blue, green and red correspond the masses of the $Z^{\prime}$ boson 3\,Tev, 4\,TeV and 5\,TeV respectively.}
\label{fig2}
\end{figure}

Let us first focus on that same-sign di-muon process. With these branching ratios in hand, we may analyze the behavior of the cross section 
$\sigma (p\,p \rightarrow Z^{\prime} \rightarrow S_3\,S_3 \rightarrow \ell^{\pm}\ell^{\pm}W^{\mp}W^{\mp})$ with  $m_{S_3}$ for three values of $m_{Z^{\prime}}$: 3\,TeV, 4\,TeV and 5\,TeV. The results are displayed at Figure. \eqref{fig2}. It was claimed in Ref.\cite{Cox:2017eme} that for $m_{Z^\prime}=3$\,TeV and $m_{S_3}=m_{Z^{\prime}}/4$, a cross section of $\sigma (pp \rightarrow Z^{\prime} \rightarrow S_3\,S_3 \rightarrow \ell^{\pm}\ell^{\pm}W^{\mp}W^{\mp})\simeq 0.1$\,fb is necessary for a 5$\sigma$ discovery at the LHC with 300\,fb$^{-1}$. The results in Figure. \eqref{fig2} are in accordance with the simulation done in Ref.\cite{Cox:2017eme}

\begin{figure}[ht]
\centering
\includegraphics[scale=0.5]{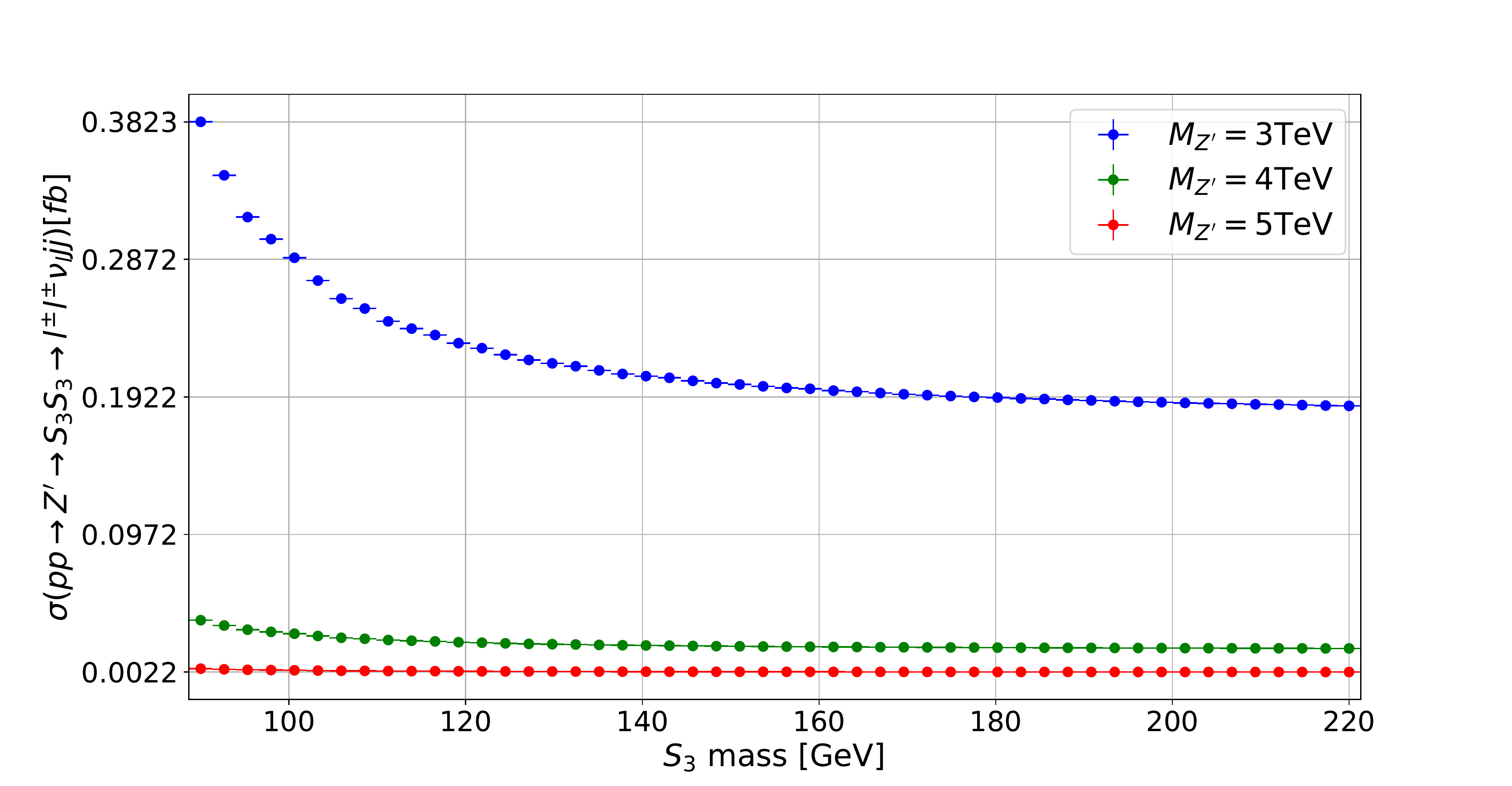}
\caption{ Behavior of the cross section $\sigma ( pp \rightarrow Z^{\prime} \rightarrow S_3\,S_3 \rightarrow \ell^{\pm}\ell^{\mp}\ell^{\mp}\nu_{\ell}\jmath\jmath)$ varying with $m_{s_3}$ at LHC with 14\,TeV. The colors blue, green and red correspond the masses of the $Z^{\prime}$ boson 3\,Tev, 4\,TeV and 5\,TeV respectively.}
\label{fig3}
\end{figure}

For the case of tri-lepton final states, the behavior of the cross section with $m_{S_3}$ is displayed at Figure. \eqref{fig3}. It was claimed in Ref.\cite{Accomando:2017qcs} that for $m_{Z^{\prime}}=4$\,TeV and $m_{s_3}=400$\,GeV, a cross section $\sigma ( pp \rightarrow Z^{\prime} \rightarrow S_3\,S_3 \rightarrow \ell^{\pm}\ell^{\mp}\ell^{\mp}\nu_{\ell}\jmath\jmath)\simeq 0.37$\,fb is necessary for the LHC obtains a signal-to-background ratio of $S/\sqrt{B} \simeq 10$ with 300\,fb$^{-1}$. Although our results in Figure. \eqref{fig3} is a little smaller than the claimed by the simulation, however they agree in order of magnitude.


According to the Figures. \eqref{fig2} and \eqref{fig3}, it seems that the process $pp \rightarrow Z^{\prime} \rightarrow S_3\,S_3 \rightarrow \ell^{\pm}\ell^{\pm}W^{\mp}W^{\mp} $ appears to be the better channel to probe the RHNs within the  331$\nu_R$ model. 

\begin{figure}[ht]
\centering
\includegraphics[scale=0.7]{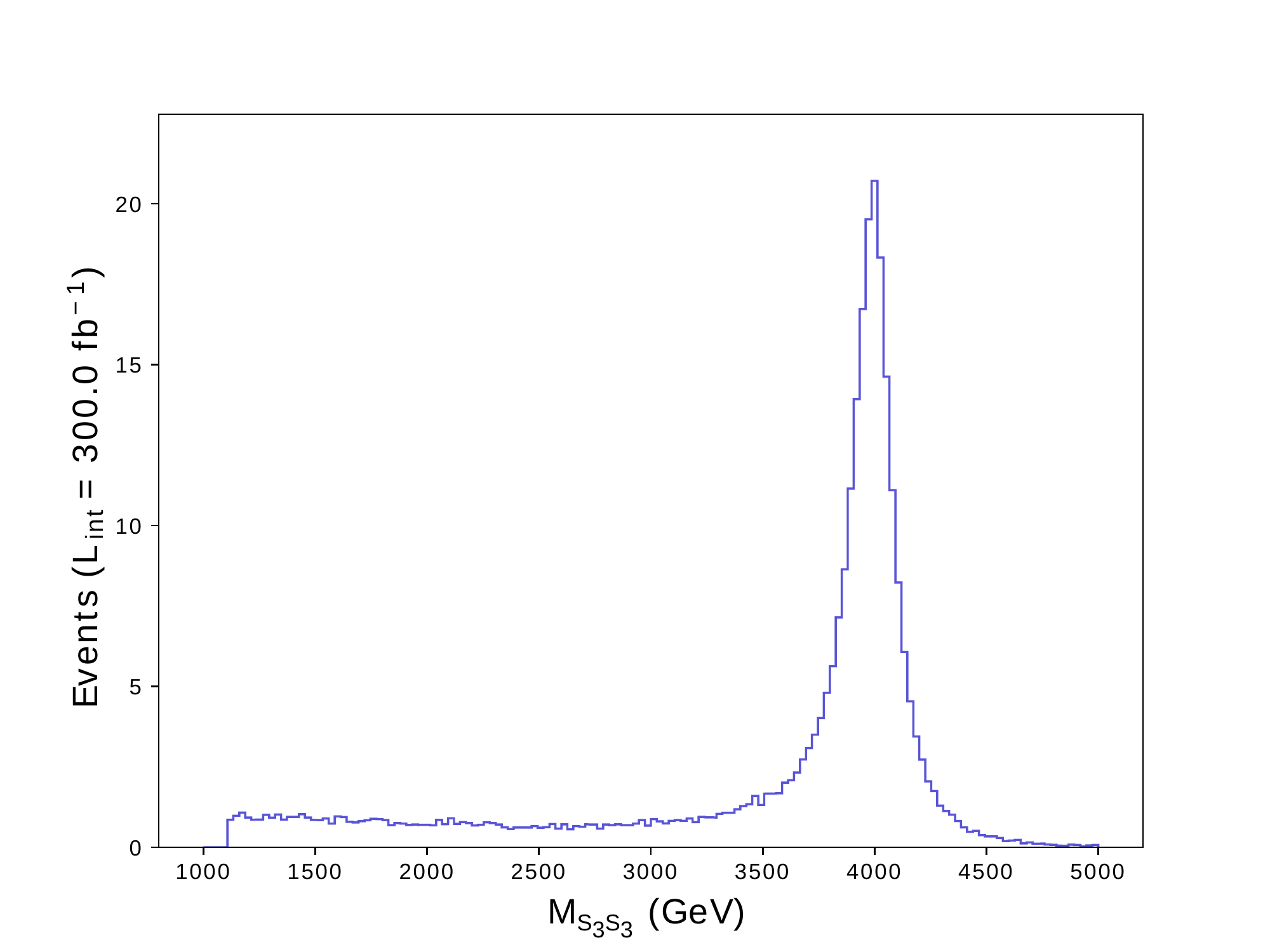}
\caption{ Process $p\,p\rightarrow Z^{\prime} \rightarrow S_3\,S_3$ with $m_{Z^{\prime}}= 4$\,TeV, $m_{S_3}=200$\,GeV and $300\,\text{fb}^{-1}$ luminosity.}
\label{fig5}
\end{figure}

We finish this section calculating the number of events predicted by the process $pp\rightarrow Z^{\prime} \rightarrow S_3 S_3$ at the LHC with 14\,TeV and 300\,fb$^{-1}$ of luminosity and $m_{Z^\prime}=4$\,TeV and $m_{S_3}=200$\,GeV. The result is displayed at Figure. \eqref{fig5}. Perceive that at the resonance of $Z^{\prime}$ we have a considerable  enhancement of the number of event. This is a very promising prediction of the model. This result shows the relevance of searching RHNs at the resonance of $Z^{\prime}$.
%
%
%
%
%
%

\section{Conclusions}

Right-handed neutrinos is an undoubted signature of type-I seesaw mechanism. Search for their signature  is a very well motivated topic in particle physics. The benchmark framework of this mechanism is the $B-L$ model.  

It is very hard to detect RHNs since it mix very weakly with the standard neutrinos. However, it was recently perceived that the discovery of this particle may depend strongly of the discovery of a neutral gauge boson, $Z^{\prime}$. In this case RHNs production at the LHC may be considerably enhanced if its production occurs at the resonance of $Z^{\prime}$ through the process $p\,p \rightarrow Z^{\prime} \rightarrow S\,S$. The main signature of these RHNs are trilepton final states, $\sigma ( pp \rightarrow Z^{\prime} \rightarrow S_3\,S_3 \rightarrow \ell^{\pm}\ell^{\mp}\ell^{\mp}\nu_{\ell}\jmath\jmath)$ , or same-sign  di-muon and a boost di-boson, $pp \rightarrow Z^{\prime} \rightarrow S_3\,S_3 \rightarrow \ell^{\pm}\ell^{\pm}W^{\mp}W^{\mp} $. Recent simulation studies done in the framework of $B-L$ model have conclude that RHNs may be discovered at the LHC if these cross sections are of order of $10^{-1}$\,fb. 

We analyzed such issue within the 331$\nu_R$ model. Our results are optimistic once the model provides  $\frac{\Gamma(Z^{\prime} \rightarrow S_3\,S_3)}{\Gamma(Z^{\prime} \rightarrow \ell^+ \ell^-)} \simeq 4.7$. Thus we studied the processes  $p\,p \rightarrow Z^{\prime} \rightarrow S_3\,S_3 \rightarrow \ell^{\pm}\ell^{\pm}W^{\mp}W^{\mp}$ and $ pp \rightarrow Z^{\prime} \rightarrow S_3\,S_3 \rightarrow \ell^{\pm}\ell^{\mp}\ell^{\mp}\nu_{\ell}\jmath\jmath $. We found that the cross sections for both processes  take values around $10^{-1}$\,fb for RHN mass around 200\,GeV and $m_{Z^{\prime}} \simeq 4$\,TeV. These results suggest that the 331$\nu_R$ may be an interesting  benchmark model for simulation studies probing signature of RHNs at future LHC runs.

\section{Appendix}

Within the 331$\nu_R $ model the natural choice for $v_\Phi$ and $v_\sigma$ are MeV and TeV, respectively. This leads to $v_\Delta$ at eV scale. In order to provide neutrino mass spectrum that explain solar and atmospheric neutrino oscillations, $v_\Delta$ at eV scale requires Yukawa couplings, $G_{ab}$, in the range $\approx 10^{−3} − 10^{−2}$  As a realistic choice for the set of Yukawa couplings, we take: $G_{11} = −0.0022$, $G_{12} = G_{21} = 0.0012$, $G_{13} = G_{31} = 0.0049$, $G_{22} = −0.0257$, $G_{23} = G_{32}= −0.0169$, $G_{33} = -0.0189$. With this set of Yukawa coupling values, the diagonalization of $M_{\nu}$ in Eq. (\ref{seesaw}) provides the following  masses for the standard neutrinos:
\begin{eqnarray}
m_{N_1}&\approx& 0,  \nonumber\\
m_{N_2}&=& 8.6 \times 10^{-3}\,\text{eV},  \\
m_{N_3}&=& 5.02\times 10^{-2}\,\text{eV},  \nonumber 
\end{eqnarray}
which yield the following mass differences
\begin{eqnarray}
(\Delta m_{21})^{2} &=& 7.50 \times 10^{-5}\,\text{eV}^{2} \quad  (\textit{solars})\nonumber\\
(\Delta m_{32})^{2} &=& 2.524 \times 10^{-3}\,\text{eV}^{2} \quad  (\textit{atmospheric}),
\end{eqnarray}
which explain solar and atmospheric neutrino oscillation experimental results \cite{Patrignani:2016xqp}.

This same set of Yukawa couplings define, also, the texture of the mass matrix of the heavy neutrinos given in Eq. (\ref{seesaw}).  Diagonalizing that mass matrix we obtain:
\begin{eqnarray}
m_{S_1}&=& 4.2328 \,\, \,\text{MeV}, \nonumber\\
m_{S_2}&=& 34.4379 \,\,  \text{GeV}, \\
m_{S_3}&=& 199.873\,\,  \text{GeV}. \nonumber
\end{eqnarray}

The respective rotating matrix are
\begin{equation}
U_{PMNS}=\left(\begin{array}{ccc}
  0.830   &  0.540   &   -0.120    \\
\newline \\
  -0.250  &  0.590   &   0.720      \\
\newline \\
  0.440   &  -0.600  &   0.690     \\
\end{array}\right),\quad
U_{R}=\left(\begin{array}{ccc}
 0.854     &    -0.509   &    0.107 \\
\newline \\
 -0.257    &     -0.592     &     -0.764 \\
\newline \\
 0.453      &    0.625    &     -0.636 \\
\end{array}\right).
\end{equation}
In our analysis of $S_3$ production we used $U_R$ given above.

\bibliographystyle{JHEPfixed}
\bibliography{sample}

\end{document}